\documentclass{article}

\usepackage{amsmath}
\usepackage{amssymb}
\usepackage{graphicx}
\usepackage{multirow}
\usepackage{natbib}
\usepackage{hyperref}
\usepackage{authblk}

\usepackage{geometry}
\def \psia {\varphi}

\def \psii {\psia_1}
\def \psij {\psia_2}

\def \xi    {x_1}

\def \yi    {y_1}
\def \ybi   {\overline{y}_1}
\def \yj    {y_2}
\def \ybj   {\overline{y}_2}
\def \yk    {y_k}
\def \ybk   {\overline{y}_k}

\def \Ka    {\mathcal{K}_a}
\def \Kb    {\mathcal{K}_b}
\def \Oc    {\mathcal{O}_a}
\def \Od    {\mathcal{O}_b}

\def \Sa    {\mathcal{S}_a}
\def \Se    {\mathcal{S}_b}
\def \Sf    {\mathcal{S}_c}
\def \Sg    {\mathcal{S}_7}
\def \Sg    {\mathcal{S}_d}
\def \Rd    {\mathcal{R}_a}
\def \Re    {\mathcal{R}_b}
\def \Rf    {\mathcal{R}_c}

\def \Ham {\mathcal{H}}
\def \avHam {\bar \Ham}
\def \HamS {\Ham_0}

\def \inc {I}
\def \ep {{\rm e}}
\def \ii {{\rm i}}
\def \degree {^{\circ}}

\def \Ru {R}

\def \DeltaIn {D}

\title{\textbf{Stability maps for the 5/3 mean motion resonance between Ariel and Umbriel with inclination}}

\author{Sérgio R. A. Gomes$^1$ and Alexandre C. M. Correia$^{1,2}$}

\date{$^1$ \small CFisUC, Departamento de F\'{i}sica, Universidade de Coimbra, 3004-516 Coimbra, Portugal \\
$^2$ IMCCE, Observatoire de Paris, PSL Universit\'{e}, 77 Av. Denfert-Rochereau, 75014 Paris, France}
\begin{document}
	\maketitle
	
\begin{abstract}
    The evolution of five the largest satellites of Uranus during the cross of the 5/3 MMR between Ariel and Umbriel is strongly affected by chaotic motion. Studies with numerical integrations of the equations of motion and analysis of Poincaré surfaces provided helpful insights to the role of chaos on the  system. However, they lack a quantification of the chaos in the phase-space. We constructed stability maps using the frequency analysis method. We determined that for lower energies, the phase-space is mainly stable. As the energy increases, the chaotic regions replace the stable motion, until only small, localized  libration areas remain stable.
\end{abstract}

\section{Introduction}
The dynamical evolution that led to the current configuration of the regular satellites of Uranus remains an enigma \citep[e.g.][]{Pollack_etal_1991,Szulagyi_etal_2018, Ishizawa_etal_2019,Ida_etal_2020,Rufu_Canup_2022}. 
The current large eccentricities ($\sim 10^{-3}$) cannot be explained by mutual interactions between the satellites \citep[][]{Squyres_etal_1985,Smith_etal_1986,Peale_1988}. Additionally, the surfaces of Miranda, Ariel, and Titania display evidences of  large scale surface melting \citep[][]{Dermott_etal_1988,Peale_1988,Tittemore_1990}.
Tidal interactions raised on the planet induce a differential outward motion of the satellites \citep[eg.][]{Peale_1988, Tittemore_Wisdom_1988, Tittemore_Wisdom_1989, Tittemore_Wisdom_1990, Pollack_etal_1991, Cuk_etal_2020}. This migration likely resulted in multiple encounters with mean motion resonances (MMRs) over their evolution. Although currently we do not observe any MMR in the system, tidal models suggests that the 5/3 MMR between Ariel and Umbriel was the latest to be crossed \citep[eg.][]{Peale_1988, Tittemore_Wisdom_1988,Cuk_etal_2020,Gomes_Correia_2023}, possibly exciting the eccentricities and inclination of the moons. 

\citet{Tittemore_Wisdom_1988} carried a detailed study on the passage through the 5/3 Ariel-Umbriel MMR in the planar approximation and for small eccentricities. Their results shown that chaotic motion has a significant impact on the dynamical evolution during the resonance crossing. In fact, this chaos can drive the eccentricities of Ariel and Umbriel to much higher values than the initial ones and provides a mechanism to break the resonance.

Using a $N$-body integrator, \citet{Cuk_etal_2020} studied the passage through the 5/3 Ariel-Umbriel MMR. The authors account for the five regular Uranian moons and assessed the role of the eccentricity,  the inclination, and the spin for the outcome of the resonance crossing. Their results show that, during resonant passage, the eccentricities and the inclinations of the five Uranian satellites are excited by chaotic motion,  even if they are not in resonance. 

\citet{Gomes_Correia_2023} also revisited the intricate 5/3 MMR, in order to understand the role of inclination in the outcome of the passage. They performed a similar analysis as the one by \citet{Tittemore_Wisdom_1988,Tittemore_Wisdom_1989}, with a two-body secular model in the circular approximation with low inclinations, and a more robust tidal model. Resorting to the Poincaré surface section method, their analysis confirms the results obtained by \citet{Tittemore_Wisdom_1988} and \citet{Cuk_etal_2020}, as well as the theoretical predictions from \citet{Dermott_etal_1988}, that chaotic motion rules the dynamics of the MMR between Ariel and Umbriel.
In this work, we re-evaluate the stability of the 5/3 MMR, but now applying the frequency analysis method \citep{Laskar_1990, Laskar_1993PD} to access the stability of the system.

\section{Model}\label{sec:Model}

To easily analyse the stability of the 5/3 MMR, we need to resort to a simplified model, with a reduced number of degrees of freedom. For that, we adopt the secular resonant two-satellite circular model with low inclinations developed in \citet{Gomes_Correia_2023}.

The model considers an oblate central body of mass $m_0$ (Uranus) surrounded by two point-mass bodies $m_1$, $m_2 \ll m_0$ (satellites), where the subscript 1 refers to the inner orbit (Ariel) and the subscript 2 refers to the outer orbit (Umbriel) and departs from a Hamiltonian truncated to the first order in the mass ratios, $m_k/m_0$, zeroth order in the eccentricities, and second order in the inclinations, $\inc_k$ (with respect to the equatorial plane of the central body). After the high frequency angles are averaged, the Hamiltonian reads as \begin{equation}
\label{cartHam}
    \begin{split}
       \avHam&=\left(\Ka+\Sa\right)(\yi\ybi+\yj\ybj)+\Kb(\yi\ybi+\yj\ybj)^2\\
       &+\left(\Oc+\Se\right) \yi\ybi+\left(\Od+\Sf\right) \yj\ybj+\frac{\Sg}{2}(\yi\ybj+\ybi\yj) \\
       &+\frac{\Rd}{2}(\yi^2+\ybi^2)+\frac{\Re}{2}(\yj^2+\ybj^2)+\frac{\Rf}{2}(\yi\yj+\ybi\ybj) \ ,
    \end{split}
\end{equation}
where $K$ stands for the Keplerian coefficients, $O$ for the oblateness coefficients, $S$ for secular the coefficients, and $R$ for the resonant coefficients \citep[see appendix A in][]{Gomes_Correia_2023}. The Hamiltonian is written as a function of a set of canonical complex rectangular coordinates ($\yi,\ii \, \ybi,\yj,\ii \, \ybj$), given by
\begin{equation}\label{eq:complex_cartesian_coordinates}
    \yk = \sqrt{\beta_k\sqrt{\mu_k a_k}\,(1-\cos \inc_k)}\,\ep^{\ii\psia_k} \, \approx  \, \inc_k \sqrt{\frac{\Gamma_k}{2}} \ep^{\ii\psia_k}\ ,
\end{equation}
where $\ybk$ is the complex conjugate of $\yk$, $a_k$ is the semi-major axis, $\beta_k=m_0m_k/(m_0+m_k)$, $\mu_k=G(m_0+m_k)$, $G$ is the gravitational constant, and
	$\psia_k = \frac{5}{2}\lambda_2-\frac{3}{2}\lambda_1 - \Omega_k$ 
are the resonant angles, where $\lambda_k$ is the mean longitude and $\Omega_k$ is the longitude of the ascending node. 
\begin{equation}\label{Gb}
\Gamma_1  = \frac{3}{2} \Gamma \left( 1 - \Delta \right) \quad \text{and} \quad \Gamma_2 = -\frac{3}{2} \Gamma \left(1 -\frac{5}{3} \Delta \right)  
\end{equation}
are constant parameters, where ${\Delta=(\beta_1\sqrt{\mu_1a_1} \cos I_1 +\beta_2\sqrt{\mu_2a_2} \cos I_2})/\Gamma$ is the normalised total orbital angular momentum, 
with $\Gamma=2.6684 \times 10^{-12}$~$M_\odot \, \mathrm{au}^2 \, \mathrm{yr}^{-1}$.
The conservative equations of motion are simply obtained from the Hamilton equations, yielding
\begin{equation}\label{eq:conservative_motion_equations1}
    \dot{y}_{1}=\ii  \Bigg[\left(\Ka+\Sa\right)\yi+2\Kb\left(\yi\ybi+\yj\ybj\right)\yi
    +\left(\Oc+\Se\right)\yi+\frac{\Sg}{2}\yj+\Rd\ybi+\frac{\Rf}{2}\ybj\Bigg] 
\end{equation}   
and 
\begin{equation}\label{eq:conservative_motion_equations2}
    \dot{y}_{2}=\ii  \Bigg[\left(\Ka+\Sa\right)\yj+2\Kb\left(\yi\ybi+\yj\ybj\right)\yj
    +\left(\Od+\Sf\right)\yj+\frac{\Sg}{2}\yi+\Re\ybj+\frac{\Rf}{2}\ybi \Bigg] \ .
\end{equation}

The dynamics of the 5/3~MMR essentially depends on $\Delta$. We introduce the quantity
\begin{equation}\label{delta:equation}
\delta = \frac{\Delta}{\Delta_r} - 1 \quad \text{with} \quad \Delta_r=0.7681  \ ,
\end{equation}
which measures the proximity to the nominal resonance.

\section{Stability maps}

\citet{Gomes_Correia_2023} analysed the behaviour of Ariel and Umbriel at different stages of the crossing of the 5/3 MMR resorting to Poincaré surface sections. Here, we revisit the global dynamics of this resonance using stability maps. 
To this end, we adopt the frequency analysis method \citep{Laskar_1990, Laskar_1993PD} to map the diffusion of the orbits.

For coherence, we fix $\delta=-2 \times 10^{-6}$ (Eq.\,(\ref{delta:equation})) as in \citet{Gomes_Correia_2023}, for which the most diverse dynamics can be found, and adopt the physical properties of the system from Table~\ref{table:physical_orbital_parameters}.
Since the Hamiltonian (Eq.\,(\ref{cartHam})) is a four-degree function of $\yk$, the intersection of the constant energy manifold by a plane may have up to four roots (families).
Each family corresponds to a different dynamical behaviour, and so we must plot one of them at a time.
However, the families are symmetric, and actually we only need to show two of them.
We chose to represent the families with the positive roots (that we dub 1 and 2).

For each energy value, we build a grid of $200 \times 200$ equally distributed initial conditions in the plane ($y_{1,i},y_{1,r}$), where $y_{k,i}$ and $y_{k,r}$ correspond to the imaginary and real parts of $y_k$, respectively. We fix $y_{2,r}=0$ for all initial conditions and compute $y_{2,i}$ for each family from the total energy (Eq.\,(\ref{cartHam})).
We then numerically integrate the equations of motion (\ref{eq:conservative_motion_equations1}) and (\ref{eq:conservative_motion_equations2}) for a time $T$.
Finally, we perform a frequency analysis of $\yi$, using the software TRIP \citep{Gastineau_Laskar_2011} over the time intervals $[0,T/2]$ and $[T/2,T]$, and determine the main frequency in each interval, $f_{\text{in}}$ and $f_{\text{out}}$, respectively.
The stability of the orbit is measured by the index
\begin{equation}
\DeltaIn \equiv \left\lvert1 - \frac{f_{\text{out}}}{f_{\text{in}}}\right\rvert \ ,
\label{deltaindex}
\end{equation}
which estimates the stability of the orbital long-distance diffusion \citep{Dumas_Laskar_1993}.
The larger $\DeltaIn$, the more orbital diffusion exists.
For stable motion, we have $\DeltaIn \sim 0$, while $\DeltaIn \ll 1$ if the motion is weakly perturbed, and $\DeltaIn \sim 1$ when the motion is irregular.
It is difficult to determine the precise value of $\DeltaIn$ for which the motion is stable or unstable, but a threshold of stability $\DeltaIn_s$ can be estimated such that most of the trajectories with $\DeltaIn < \DeltaIn_s$ are stable \citep[for more details see][]{Couetdic_etal_2010}.

The diffusion index depends on the considered time interval.
Here, we integrate the equations of motion for $T=10^4$~yr, because this interval is able to capture the main characteristics of the dynamics regarding the resonant frequency, which lies within the range $\sim 60$~yr.
With this time interval, we estimate that $\DeltaIn_s \sim 10^{-4}$.
The diffusion index $\DeltaIn$ is represented by a logarithmic colour scale calibrated such that blue and green correspond to stable trajectories ($\DeltaIn \ll \DeltaIn_s$), while orange and red correspond to chaotic motion ($\DeltaIn \gg \DeltaIn_s$).

\begin{table}
\caption{Present physical and orbital properties of the Uranian system \citep{Thomas_1988, Jacobson_2014}.}
\label{table:physical_orbital_parameters}      
\centering          
\begin{tabular}{c c c c c c }     
	\hline\hline       
    &$m$ (M\textsubscript{\(\odot\)}$\times10^{-10}$)	& $\Ru$ (km) & $\langle T_{\rm rot}\rangle$ (day)& $J_2 (\times10^{-3})$ & $C/(m_0R^2)$ \\
    \hline
    Uranus & 436562.8821 & 25559. & 0.7183 & 3.5107 & 0.2296\\
	\hline\hline       
	Satellite 	& $m$ (M\textsubscript{\(\odot\)}$\times10^{-10}$)	& $\Ru$ (km) & $\langle T_{\rm orb}\rangle$ (day) & $\langle a \rangle$ ($\Ru$) & $\langle \inc \rangle$ ($\degree$)\\
	\hline                    
	Ariel 	& 6.291561 	& 578.9  & 2.479971 	& 7.468180 	 & 0.0167\\
	Umbriel & 6.412118 	& 584.7  & 4.133904 	& 10.403550	 & 0.0796\\
    \hline
\end{tabular}
\end{table}

\begin{figure}
    \centering
    \includegraphics[width=\textwidth]{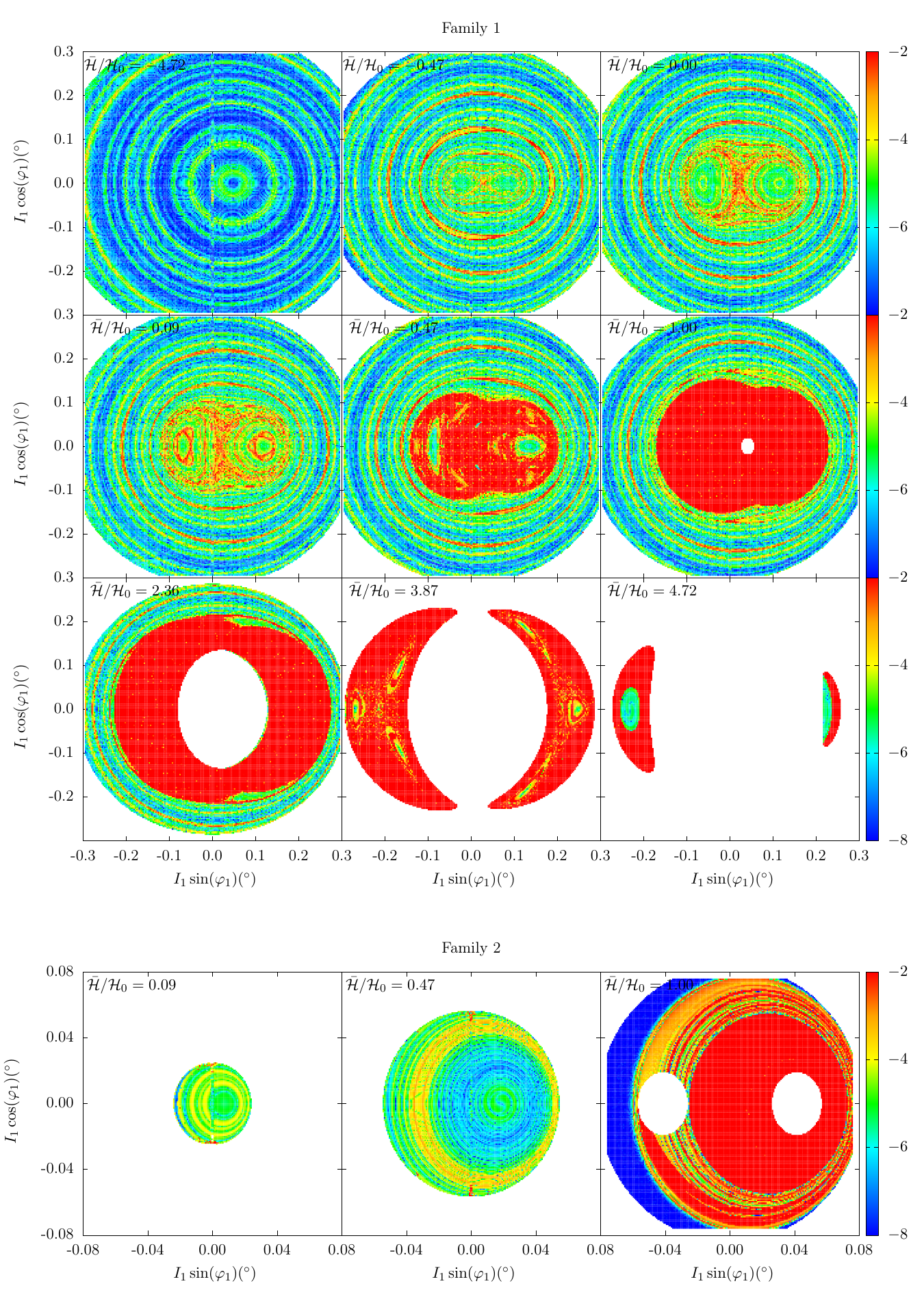}
    \caption{Stability maps for Ariel in the plane ($\inc_1 \sin \psii, \inc_1 \cos \psii $) with $\cos \psij=0$ and $\delta=-2 \times 10^{-6}$. The colour scale corresponds to the relative frequency diffusion index in logarithmic scale (Eq.\,(\ref{deltaindex})). 
    Each panel was obtained with a different energy value and $\Ham_0=1.06 \times 10^{-19}$~$M_\odot \, \mathrm{au}^2 \, \mathrm{yr}^{-2}$.
    \label{fig:stabmap}}
\end{figure}

In Fig.~\ref{fig:stabmap}, we show the stability maps for Ariel. 
We rescale $\yk$ by $\sqrt{\Gamma_k/2}$, and so we actually plot the maps in the plane ($\inc_1 \sin \psii, \inc_1 \cos \psii $) with $\cos \psij = 0$ (Eq.\,(\ref{eq:complex_cartesian_coordinates})).
Each panel corresponds to a different energy value $\avHam/\Ham_0$, where $\Ham_0=1.06 \times 10^{-19}$~$M_\odot \, \mathrm{au}^2 \, \mathrm{yr}^{-2}$ is the energy of the transition between the circulation and libration regions, i.e., the energy of the separatrix.
The lowest energies occur in the circulation regions, $\avHam<\HamS$, while the largest energies occur in the libration region, $\avHam>\HamS$.
The inner circulation region is delimited by $0<\avHam<\HamS$, where $\avHam=0$ corresponds to the energy of the equilibrium point with $y_1=y_2=0$.
For this energy range, there are four families, while for the remaining energies only two families exist.

For $\avHam \ll 0$ (Fig.~\ref{fig:stabmap}\,a), only family 1 exists, and we observe that the system is always stable, corresponding to trajectories in the outer circulation region.
As the energy increases, two islands appear, corresponding to trajectories that are in the libration region (in resonance).
Initially, the motion in these new regions is also stable, but the sepatrix and some localized concentric regions outside the separatrix are chaotic.
As the energy approaches the threshold $\avHam = 0$ (Figs.~\ref{fig:stabmap}\,b,c), the chaotic regions expand for the vicinities of the separatrix. 
For  $0<\avHam<\HamS$ (family 1), the chaotic regions increase even further, while the resonant islands shrink (Figs.~\ref{fig:stabmap}\,d,e), until they completely disappear for $\avHam=\HamS$ (Fig.~\ref{fig:stabmap}\,f). Note that up to these energies, outside the chaotic regions, the circulation region remains stable.
For this specific energy range, we also need to plot family 2.
Close to $\avHam = 0$, we observe stable motion in the inner circulation region (Figs.~\ref{fig:stabmap}\,j,k). However, as we approach $\avHam=\HamS$, this area is replaced by a chaotic region (Fig.~\ref{fig:stabmap}\,l).
Finally, for $\avHam>\HamS$, we observe that the stable region progressively vanishes, and chaotic motion dominates the phase-space, where only small libration regions remain stable (Figs.~\ref{fig:stabmap}\,g,h,i).
In this energy range, we only have family 1 and trajectories in the outer circulation region also do not exist.
Moreover, there is also a forbidden region at the centre of each panel that grows with the energy value, while the libration areas shrink.
The results in Fig.~\ref{fig:stabmap} are in perfect agreement with those shown in Fig.~3 from \citet{Gomes_Correia_2023} using Poincar\'e surface sections.

\section{Conclusion}

As observed by \citet{Gomes_Correia_2023} and several previous works \citep{Dermott_etal_1988,Tittemore_Wisdom_1988,Cuk_etal_2020}, chaos has a strong presence on the passage through the 5/3 MMR between Ariel and Umbriel. However, the analysis of this resonance with stability maps allows to quantify the chaos for each region.

The dynamics of the 5/3~MMR between Ariel and Umbriel is very rich and depends on the energy of the system.
In fact, the energy depends on the value of the inclinations (Eq.\,(\ref{cartHam})), given by the variables $y_1$ and $y_2$ (Eq.\,(\ref{eq:complex_cartesian_coordinates})).
Therefore, the value of the inclinations of Ariel and Umbriel when the system encounters the resonance can trigger completely different behaviours.
For $\avHam < 0$, the phase-space is dominated by stable orbits. As we approach the sepatrix energy, $\avHam \sim \HamS$, the chaotic motion engorges the low inclination regions, while the outer circulation regions remain stable.
Finally, for $\avHam \gg \HamS$, only small libration regions remain stable, surrounded by large chaotic areas. This is a new result, since surface sections appeared to have exclusive quasi-periodic motion for energies near the equilibrium resonance points.

The dynamical analysis with stability maps can be extended to several combinations of the variables. Indeed, we do not need to choose a specific projection plane, and so the choice of the phase-space plane is much less restricted than for the Poincaré surface sections \citep[see also][]{Alves-Carmo_etal_2023}. Allied to the quantification of chaos, the stability maps method thus provides a more exhaustive analysis of the dynamics of the resonance passage.

\section*{Acknowledgements}
This work was supported by grant
SFRH/BD/143371/2019,
and by projects
CFisUC (UIDB/04564/2020 and UIDP/04564/2020),
GRAVITY (PTDC/FIS-AST/7002/2020), 
PHOBOS (POCI-01-0145-FEDER-029932), and
ENGAGE SKA (POCI-01-0145-FEDER-022217),
funded by COMPETE 2020 and FCT, Portugal.
We acknowledge the Laboratory for Advanced Computing at University of Coimbra (\href{https://www.uc.pt/lca}{https://www.uc.pt/lca}) for providing the resources to perform the stability maps with high resolution.

\bibliography{bibliography.bib}
\bibliographystyle{apalike}
\end{document}